\documentclass[twocolumn,showpacs,preprintnumbers,amsmath,amssymb,aps,prb,superscriptaddress]{revtex4-1}

\pdfoutput=1
\usepackage[pdftex]{graphicx}
\usepackage[english]{babel}
\usepackage{soul}
\usepackage{cleveref}
\usepackage{bbold}
\usepackage{mathtools}
\usepackage{multirow}
\usepackage{colortbl}
\definecolor{kugray5}{RGB}{224,224,224}

\usepackage{dcolumn}
\usepackage{bm}

\usepackage{color}
\usepackage{amstext}
\usepackage{braket}
\usepackage{mathdots}
\usepackage{tikz}
\usetikzlibrary{matrix,decorations.pathreplacing}

\begin{document}


\title{Generalization of Zak's phase for lattice models with non-centered  inversion symmetry axis}

\author{A. M. Marques}
\email{anselmomagalhaes@ua.pt}
\affiliation{Department of Physics $\&$ I3N, University of Aveiro, 3810-193 Aveiro, Portugal}
\author{R. G. Dias}
\affiliation{Department of Physics $\&$ I3N, University of Aveiro, 3810-193 Aveiro, Portugal}

\date{\today}


\begin{abstract}
We show how the presence of inversion symmetry in a one-dimensional (1D) lattice model is not a sufficient condition for a quantized Zak's phase. This is only the case when the inversion axis is at the center of the unit cell. When the inversion axis is not at the center, the modified inversion operator within the unit cell gains a $k$-dependence in some of its matrix elements which adds a correction term to the usual Zak's phase expression \cite{Zak1989}, making it in general deviate from its quantized value. A general expression that recovers a quantized Zak's phase in a lattice model with a unit cell of arbitrary size and arbitrarily positioned inversion axis is provided in this paper, which relates the  quantized value with the eigenvalues of a modified parity operator at the inversion invariant momenta.
\end{abstract}


\maketitle
%
%

The topological characterization of 1D lattice models relies on the calculation of the sum of the so-called Zak's phase \cite{Zak1989} $\gamma$ over all occupied bands.
(Non-)trivial models yield $\gamma=(\pi) 0 \, \text{mod} (2 \pi)$.
These values of $\gamma$ can be confirmed by calculating the Wannier centers for each band and also by looking into the parity of the states at the middle and edge of the bands.
Essential to this latter argument is the fact that the inversion operation on Bloch states of a 1D model, in what concerns the action within the unit cell, can be represented by an unitary operator $\hat \pi$ which is independent of the unit cell and therefore is independent of the Bloch state momentum $k$ \cite{Asboth2016}.

Here we show that, in 1D models with $k$-dependent inversion ($\mathcal{I}$) symmetry within the unit cell, a correction term has to be added to the Zak's phase in order to preserve its quantization. This follows from the fact that the inversion operator can not be written as $\hat \Pi \ket{v_j(k)}=\hat \Pi \big(\ket{k} \otimes \ket{u_j(k)}\big) = \vert -k \rangle \otimes {\hat \pi}\ket{u_j(k)}$ with $\hat \pi$ independent of $k$, where $\ket{v_j(k)}$ is the eigenvector of band $j$ of a periodic chain, factorized as the product of a Bloch plain wave $\ket{k}$ and the eigenstate $\ket{u_j(k)}$ of the $k$-space bulk Hamiltonian. More specifically, in these models, it is never possible to choose a unit cell such that the inversion symmetry maps  each unit cell onto a single unit cell, as is usually assumed \cite{Alexandradinata2014,Miert2017}.

\begin{figure}[h]
\begin{center}
\includegraphics[width=0.47 \textwidth]{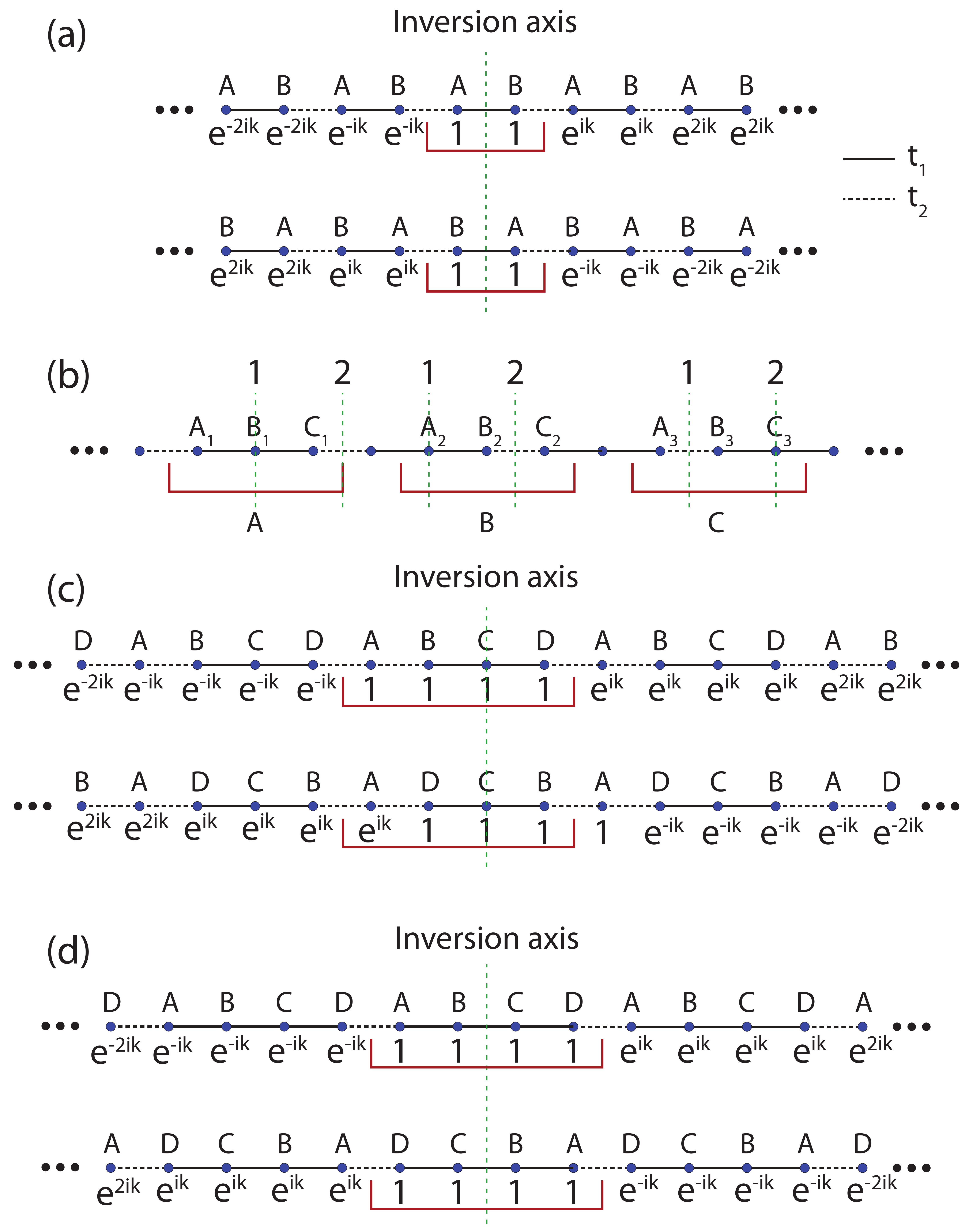}
\end{center}
\caption{(a) Scheme of an inversion operation in a periodic SSH chain. Both sites in a given unit cell have the same phase, before and after the inversion.
(b) Possible choices of unit cell and inversion axes for a periodic $t_1t_1t_2$ chain. Only unit cell A has an inversion axis at its center.
(c) Scheme of an inversion operation in a periodic $t_1t_1t_2t_2$ chain. After the inversion is performed, the A site in each unit cell gains an extra phase factor of $e^{ik}$, relative to the other sites, as a consequence of the inversion axis being  to the right of the center. For this model, no choice of unit cell has a centered inversion axis.
(d) Scheme of an inversion operation in a periodic $t_1t_1t_1t_2$ chain. This model is the same as (c) but with one of the $t_2$ hoppings switched to $t_1$. The represented choice of unit cell has now a centered inversion axis.}
\label{fig:invsym}
\end{figure}

A convenient way to compute the Zak's phase in band $j$ is through the Wilson loop,
\begin{equation}
\mathcal{W}_{j}=\prod_{n=0}^{N-1}\braket{u_j(-\pi+ n\Delta k)\vert u_j(-\pi+(n+1)\Delta k)},
\label{eq:wilson}
\end{equation}
where we have set the momentum increment to $\Delta k=\frac{2\pi}{N}$, with $N$ the number of sites in the periodic chain.
In the continuous limit ($N\to \infty$, $\Delta k\to 0$), the Zak's phase of band $j$ becomes
\begin{eqnarray}
\gamma_j&=&i\int_{-\pi}^\pi dk \bra{u_j(k)}d_k \ket{u_j(k)}
\label{eq:usualzak}
\\
&=&\mbox{Arg}\big(\lim_{N\to \infty} \mathcal{W}_{j}\big)
\label{eq:zakarg}
\\
&=&\sum_{n=0}^{\infty} \delta\phi_n ,
\end{eqnarray}
where $d_k$ is the $k$ derivative and $ \delta\phi_n$ is the phase of element $n$ in the Wilson loop of (\ref{eq:wilson}).
The above expressions are quantized as $\gamma_j=0,\pi$, in the cases where there is a $k$-independent $\hat \pi$, such as for the Su-Schrieffer-Heeger (SSH) model \cite{Su1979} shown in Fig.~\ref{fig:invsym}(a).

The condition for a $k$-independent $\hat \pi$ is that the inversion axis is at the center of the considered unit cell.
When there are several possible unit cells with different positions for the inversion axis, as in the model of Fig.~\ref{fig:invsym}(b), one usually chooses the unit cell where this axis is centered [the first one in Fig.~\ref{fig:invsym}(b)] to compute the Zak's phase.
There are models, however, where this condition is not met for any choice of the unit cell, as demonstrated in the case of Fig.~\ref{fig:invsym}(c). 
For the unit cell considered there, the inversion operation within the unit cell, about site C, is given by
\begin{equation}
\hat \pi_k \ket{u_j(k)}
=
\begin{bmatrix}
e^{ik}&0&0&0
\\
0&0&0&1
\\
0&0&1&0
\\
0&1&0&0
\end{bmatrix}
\begin{bmatrix}
u_{j,A}(k)
\\
u_{j,B}(k)
\\
u_{j,C}(k)
\\
u_{j,D}(k)
\end{bmatrix} 
\label{eq:invmat4}.
\end{equation} 
This inversion operator is not Hermitian.
Also, in consecutive $k$ points the following relation holds, for infinitesimal $\Delta k\to dk$,
\begin{equation}
\hat \pi^\dagger_k \hat \pi_{k+dk}\simeq \mathbb{\hat 1}+idk 
\begin{bmatrix}
1&0&0&0
\\
0&0&0&0
\\
0&0&0&0
\\
0&0&0&0
\end{bmatrix} ,
\label{eq:invtransinv}
\end{equation}
where $\mathbb{\hat 1}$ is the identity operator. For models with $k$-independent $\hat \pi$, this operator is hermitian and the second term in the right hand side of (\ref{eq:invtransinv}) does not appear. 
Eigenstates with opposite momenta are related by the inversion operator within the unit cell as $\ket{u_j(-k)}=e^{i\theta_k} \hat \pi_k\ket{u_j(k)}$, for all $k\neq 0,\pi$, where $e^{i\theta_k}$ is an arbitrary phase factor that we take out for convenience from now on since they will appear as conjugate pairs in the Wilson loop \cite{Asboth2016}.
Using this last relation, together with (\ref{eq:invtransinv}), we arrive at
\begin{eqnarray}
\braket{u_j(k-dk)\vert u_j(k)}&\simeq &1+i\delta\phi,
\label{eq:ukpos}
\\
\braket{u_j(-k)\vert u_j(-k+dk)}&=&\bra{u_j(k)}\hat \pi^\dagger_k \hat \pi_{k-dk}\ket{u_j(k-dk)} \nonumber
\\
&=&\braket{u_j(k-dk)\vert u_j(k)}^* \nonumber
\\
&-& idk \braket{u_{j,A}(k)\vert u_{j,A}(k-dk)} \nonumber
\\
&\simeq &1-i\big(\delta\phi + dk\vert u_{j,A}(k)\vert^2\big),
\end{eqnarray}
for $dk<k<\pi$.
In the last step we assumed, to leading order, $ \braket{u_{j,A}(k)\vert u_{j,A}(k-dk)}\approx \vert u_{j,A}(k)\vert^2$.
However, different relations hold for the $\mathcal{I}$-symmetric momenta $\Lambda=0,\pi$, where $\ket{u_j(-\pi)}\equiv\ket{u_j(\pi)}$,
\begin{eqnarray}
\braket{u_j(-dk)\vert u_j(0)}&=&\bra{u_j(dk)}\hat \pi^\dagger_{dk}\ket{u_j(0)} \nonumber
\\
&\simeq &\bra{u_j(dk)}\hat \pi^\dagger_{0}\ket{u_j(0)} \nonumber
\\ 
&-&  idk \braket{u_{j,A}(dk)\vert u_{j,A}(0)},
\\
\braket{u_j(-\pi)\vert u_j(-\pi+dk)}&=&\bra{u_j(\pi)}\hat \pi_{\pi-dk}\ket{u_j(\pi-dk)} \nonumber
\\
&\simeq &\bra{u_j(dk)}\hat \pi_{\pi}\ket{u_j(\pi-dk)} \nonumber
\\ 
&+&  idk \braket{u_{j,A}(\pi)\vert u_{j,A}(\pi-dk)},
\label{eq:ukpi}
\end{eqnarray}
where the modified parity operators in the unit cell have the form
\begin{equation}
\hat \pi_0=
\begin{bmatrix}
1&0&0&0
\\
0&0&0&1
\\
0&0&1&0
\\
0&1&0&0
\end{bmatrix} , 
\hat \pi_\pi=
\begin{bmatrix}
-1&0&0&0
\\
0&0&0&1
\\
0&0&1&0
\\
0&1&0&0
\end{bmatrix}.
\label{eq:invopzeropi}
\end{equation}
The modified parity of the corresponding eigenstates is well defined, that is, $\hat \pi_0\ket{u_j(0)}=P_0\ket{u_j(0)}$ and $\hat \pi_\pi\ket{u_j(\pi)}=P_\pi\ket{u_j(\pi)}$, with $P_0,P_\pi=\pm 1$.

The procedure now is to substitute (\ref{eq:ukpos}-\ref{eq:ukpi}) in the computation of the Wilson loop in (\ref{eq:zakarg}) to obtain the following simplified expression for the Zak's phase,
\begin{equation}
\gamma_j\simeq\mbox{Arg}(P_0P_\pi) - \int_0^\pi dk \vert u_{j,A}(k)\vert^2,
\label{eq:zakphase}
\end{equation}
which is in general \textit{non-quantized} due to the last term.
The last term in (\ref{eq:ukpi}), with a positive sign, was disregarded as an infinitesimal surface term.
In Fig.~\ref{fig:invsym}(d), we show a similar model of a chain with four sites per unit cell (we switched one of the $t_2$ hoppings for another $t_1$) which has, as in Fig.~\ref{fig:invsym}(b), a given choice for the unit cell where the inversion axis is centered, and therefore one recovers the usual quantized Zak's phase for this choice.

\begin{figure}[h]
\begin{center}
\includegraphics[width=0.47 \textwidth]{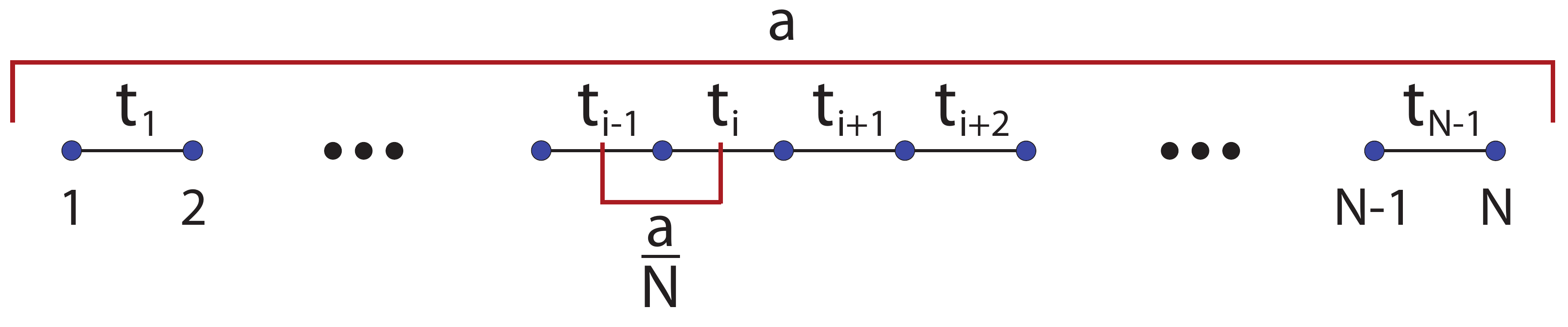}
\end{center}
\caption{Unit cell of arbitrary size $a$ and hoppings configuration. All $N$ sites are uniformly spaced, with an intersite spacing of $\frac{a}{N}$.}
\label{fig:arbitraryucell}
\end{figure}

\textit{General case}. We now generalize the previous results for the case of an arbitrary unit cell, both in size and morphology (regarding the hoppings parameters), of uniformly spaced sites with an inversion axis at an arbitrary position, as illustrated in Fig.~\ref{fig:arbitraryucell}.
The position of site $j$ is given by $r_j=(j-\frac{1}{2})\frac{a}{N}$, with $j=1,2,...,N$.
The possible positions for the inversion axis are given by $r_{m}=a(\frac{1}{2}+\frac{m}{2N})$, with $m=0,\pm 1,...,\pm N$ indicating its displacement from the center of the unit cell.
Let us first consider the usual case with the inversion axis at the center of the unit cell, that is, $m=0$. This position coincides with the central site for $N$ odd, but, for $N$ even, it corresponds to the middle point between the two consecutive sites nearest to the center of the unit cell. The inversion operator within the unit cell is $k$-independent and of the form
\begin{equation}
\hat\pi = 
\begin{bmatrix} 
 & & & &1 
\\
 & & &1 &
\\
 & &\iddots & & 
\\
 &1 & & & 
\\
1 & & & & 
\end{bmatrix},
\end{equation}
yielding the usual quantized Zak's phase.

A $k$-dependence in the inversion operator appears for $m\neq 0$. When the inversion axis is at the right of the center of the unit cell, $m>0$, the general form of the operator is given by
\begin{equation}
\hat \pi_k=
\left[
\begin{array}{c|c}
\overbrace{\begin{matrix}
& &e^{ik}
\\
 &\iddots &
\\
e^{ik} & &
\end{matrix}}^{m\times m} 
&   \\
\hline
  & 
\underbrace{\begin{matrix}
& &1
\\
 &\iddots &
\\
 1& &
\end{matrix}}_{(N-m)\times (N-m)}
\end{array}
\right].
\end{equation}
The model in Fig.~\ref{fig:invsym}(c), with the inversion operator of (\ref{eq:invmat4}), corresponds to the particular case of $N=4$ and $m=1$.
On the other hand, when the inversion axis is at the left of the center of the unit cell, $m<0$, the general form of the operator is given by
\begin{equation}
\hat \pi_k=
\left[
\begin{array}{c|c}
\overbrace{\begin{matrix}
& &1
\\
 &\iddots &
\\
 1& &
\end{matrix}}^{(N+m)\times (N+m)} 
&   \\
\hline
  & 
\underbrace{\begin{matrix}
& &e^{-ik}
\\
 &\iddots &
\\
e^{-ik} & &
\end{matrix}}_{\vert m\vert\times \vert m\vert}
\end{array}
\right].
\end{equation}
If, in the model of Fig.~\ref{fig:invsym}(c), we choose site A instead of C as the inversion center, it would correspond to $N=4$ and $m=-3$.
For any $m\neq 0$, the Zak's phase is readily obtained by a generalization of (\ref{eq:zakphase})
\begin{equation}
\gamma_j=
\begin{cases}
\mbox{Arg}(P_0P_\pi) - \sum_{s=1}^m\int_0^\pi dk \vert u_{j,s}(k)\vert^2, \text{\ \ \ \ \ \ \ \ for\ } m>0,
\\
\\
\mbox{Arg}(P_0P_\pi) + \sum_{s=0}^{\vert m\vert -1}\int_0^\pi dk \vert u_{j,N-s}(k)\vert^2, \text{\ for\ } m<0.
\end{cases}
\label{eq:generalzak}
\end{equation}

A $\pi$-quantized Zak's phase in each band, $\tilde{\gamma}_j$, can still be recovered by adding the correcting term to the usual definition of the Zak's phase [adding the sum term in (\ref{eq:generalzak}) to (\ref{eq:usualzak})] or, conversely, simply by looking at the  parity of the eigenstates at the $\mathcal{I}$-symmetric momenta [the first term in the right-hand side of (\ref{eq:generalzak})], that is, the eigenvalues of the modified $\hat{\pi}_k$ operator at those points. The expression for $\tilde{\gamma}_j$ is therefore given by
\begin{eqnarray}
\gamma_j&=&i\int_{-\pi}^\pi dk \bra{u_j(k)}d_k \ket{u_j(k)},
\\
\tilde{\gamma}_j&=&
\begin{cases}
\gamma_j + \sum_{s=1}^m\int_0^\pi dk \vert u_{j,s}(k)\vert^2, \text{\ \ \ \ \ \ \ \ for\ } m>0,
\\
\\
\gamma_j - \sum_{s=0}^{\vert m\vert -1}\int_0^\pi dk \vert u_{j,N-s}(k)\vert^2, \text{\ for\ } m<0.
\end{cases},
\label{eq:generalmodzak}
\end{eqnarray}
which agrees with
\begin{equation}
\tilde{\gamma}_j=\mbox{Arg}(P_0P_\pi),
\label{eq:paritiesmodzak}
\end{equation}
for all $m$.
\begin{table}[h]
\begin{center}
  \begin{tabular}{|c|c|c|c|c|c|c|c|c|c|c|c|}
    \hline
Unit cell&Axis&$t=\frac{t_2}{t_1}$&$\tilde{\gamma}_1$&$\tilde{\gamma}_2$&$\tilde{\gamma}_3$&$P_0^1$&$P_\pi^1$&$P_0^2$&$P_\pi^2$&$P_0^3$&$P_\pi^3$\\ 
\hline
\multirow{4}{*}{A}&\multirow{2}{*}{1}&0.5&0&0&0&+&+&-&-&+&+
\\ 
\cline{3-12}
& &2&$\pi$&0&$\pi$&+&-&+&+&-&+\\
\cline{2-12}
&\multirow{2}{*}{2}&0.5&$\pi$&$\pi$&$\pi$&+&-&-&+&+&-\\
\cline{3-12}
& &2&0&$\pi$&0&+&+&+&-&-&-\\ 
\cline{1-12}
\multirow{4}{*}{B}&\multirow{2}{*}{1}&0.5&0&0&0&+&+&-&-&+&+\\ 
\cline{3-12}
& &2&$\pi$&0&$\pi$&+&-&+&+&-&+\\ 
\cline{2-12}
&\multirow{2}{*}{2}&0.5&$\pi$&$\pi$&$\pi$&+&-&-&+&+&-\\ 
\cline{3-12}
& &2&0&$\pi$&0&+&+&+&-&-&-\\ 
\cline{1-12}
\multirow{4}{*}{C}&\multirow{2}{*}{1}&0.5&$\pi$&$\pi$&$\pi$&+&-&-&+&+&-\\
\cline{3-12}
& &2&0&$\pi$&0&+&+&+&-&-&-\\ 
\cline{2-12}
&\multirow{2}{*}{2}&0.5&0&0&0&+&+&-&-&+&+\\ 
\cline{3-12}
& &2&$\pi$&0&$\pi$&+&-&+&+&-&+\\ 
\hline
  \end{tabular}  
\end{center}
\caption{Values of the corrected Zak's phase $\tilde{\gamma}_j$ of band $j$, calculated from (\ref{eq:generalmodzak}), for each the three bands of the $t_1t_1t_2$ model of Fig.~\ref{fig:invsym}(b), considering each inversion axis of each unit cell for different $t$ dimerizations. 
Bands are ordered by decreasing energy, with 1 being the highest energy band and 3 the lowest energy band. 
The value of $P_\Lambda^j$, the parity value of the eigenstate of band $j$ at the $\mathcal{I}$-symmetric momenta $\Lambda=0,\pi$ for the modified parity operator of the correspondent axis and unit cell, is also provided. 
Signs $\pm$ stand for $\pm 1$.
The values of $\tilde{\gamma}_j$ are consistent with those calculated from (\ref{eq:paritiesmodzak}) using these tabulated modified parity values.}
\label{tab:3bands}
\end{table}
We applied these corrected expressions for the Zak's phase to the three band model of Fig.~\ref{fig:invsym}(b). 
The results are condensed in Table~\ref{tab:3bands}. 
For all cases considered there, we see a $\pi$-shift only in the Zak's phase of bands 1 and 3 when the dimerization is reversed ($t=0.5\leftrightarrow t=2$, with $t=\frac{t_2}{t_1}$). 
These are the bands that loose a state when the topological edge states appear, with symmetric energies and localized at the gaps between bands 1/3 and band 2.
On the other hand, the Zak's phase of all bands shifts $\pi$ when one considers the other inversion axis in a given unit cell for the same choice of hopping constants. This comes as a consequence of the fact that
\begin{equation}
\tilde{\gamma}_{j,2}-\tilde{\gamma}_{j,1}=\sum_{s=1}^N\int_0^\pi dk\vert u_{j,s}(k)\vert^2=\pi ,
\label{eq:axispishift}
\end{equation}
where $\ket{u_j(k)}$ is normalized and the indices $1$ and $2$ stand for the different choices of inversion axis in a unit cell, with axis $2$ at the right of axis $1$, as in Fig.~\ref{fig:invsym}(b).

The topological transition in the SSH model of Fig.~\ref{fig:invsym}(a), where the Zak's phase of both bands changes $\pi$ when crossing $t=\frac{t_2}{t_1}=1$, can also be understood from the point of view of (\ref{eq:axispishift}). 
In fact, in the SSH model, a change in the dimerization, associated with an exchange of the hopping parameters ($t_1\leftrightarrow t_2$), is formally equivalent to a change from one inversion axis to the other for the same dimerization. 
This equivalence also holds for the $t_1t_1t_2t_2$ model of Fig.~\ref{fig:invsym}(c).

\begin{table}[h]
\begin{center}
  \begin{tabular}{|c|c|c|c|c|c|c|c|c|c|c|c|c|c|}
    \hline
Model&t&$\tilde{\gamma}_1$&$\tilde{\gamma}_2$&$\tilde{\gamma}_3$&$\tilde{\gamma}_4$&$P_0^1$&$P_\pi^1$&$P_0^2$&$P_\pi^2$&$P_0^3$&$P_\pi^3$&$P_0^4$&$P_\pi^4$\\ 
\hline
\multirow{2}{*}{$t_1t_1t_2t_2$}&0.5&0&$\pi$&0&0&+&+&+&-&-&-&+&+
\\ 
\cline{2-14}
 &2&$\pi$&0&$\pi$&$\pi$&+&-&+&+&-&+&-&+\\
\cline{1-14}
\multirow{2}{*}{$t_1t_1t_1t_2$}&0.5&0&0&0&0&+&+&-&-&+&+&-&-\\
\cline{2-14}
 &2&$\pi$&0&0&$\pi$&+&-&+&+&-&-&-&+\\ 
\hline
  \end{tabular}  
\end{center}
\caption{Same as in Table \ref{tab:3bands} for the $t_1t_1t_2t_2$ and $t_1t_1t_1t_2$ models. The unit cells and inversion axes considered are those of Fig.~\ref{fig:invsym}(c) and Fig.~\ref{fig:invsym}(d), respectively.}
\label{tab:4bands}
\end{table}
The corrected Zak's phases and modified parity values at the $\mathcal{I}$-symmetric momenta for the two four bands models of Fig.~\ref{fig:invsym}(c) and Fig.~\ref{fig:invsym}(d) are shown in Table~\ref{tab:4bands}.
The presence of edge states which, similarly to the $t_1t_1t_2$ model of Table~\ref{tab:3bands}, is a consequence of the topological transition of the higher and lower energy bands (bands 1 and 4, respectively), was verified numerically for both cases.
For the inversion axis considered for the $t_1t_1t_1t_2$ model of Fig.~\ref{fig:invsym}(d) one recovers $\tilde{\gamma}_j=\gamma_j$, since the inversion axis is centered (the same goes for the A1 entry in Table~\ref{tab:3bands}). 
The $t_1t_1t_2t_2$ model, however, is different from the others consider here, in the sense that it has no unit cell with a centered inversion axis.
Therefore, the topological characterization of models such as this is determined by the modified Zak's phase $\tilde{\gamma}_j$.

\textit{Conclusion}.
We have studied 1D lattice models with inversion symmetry and have shown that, when the inversion axis is not at the center of a unit cell, the topological phase has a non-quantized Zak's phase. We identified this deviation from a $\pi$-quantized Zak's phase with a correction term that has to be included in its calculation.
This correction term comes as a consequence of the $k$-dependence of the inversion operator that acts within the unit cell.
A modified expression for the Zak's phase, yielding $\pi$-quantized values regardless of the position of the inversion axis within the unit cell, can be written simply by adding the correction term to the usual definition of the Zak's phase.
The modified expression for the Zak's phase can therefore be used as a topological invariant quantity for any 1D system with inversion symmetry.

Our results can be straightforwardly generalized to quasi-1D models (such as diamond chains) and ribbons with non-centered axes of inversion symmetry within the unit cell.


\section*{Acknowledgments}\label{sec:acknowledments}

This work is funded by FEDER funds through the COMPETE 2020 Programme and National Funds throught FCT - Portuguese Foundation for Science and Technology under the project UID/CTM/50025/2013.
RGD appreciates the support by the Beijing CSRC.
AMM acknowledges the financial support from the FCT through the grant SFRH/PD/BD/108663/2015.
We are grateful for useful discussions with E V Castro.


\bibliography{inversionsymmetry}

\end{document}